# Topological photonics by breaking the degeneracy of line node singularities in semimetal-like photonic crystals

STEFFEN BÖRM[1,3], FATEMEH DAVOODI[2], RALF KÖHL[1,3], NAHID TALEBI[2,3]

[1]*Department of Mathematics, Heinrich-Hecht-Platz 6, Kiel University, 24118 Kiel, Germany, E-mail: boerm@math.uni-kiel.de*
[2]*Institute of Experimental and Applied Physics, Kiel University, 24098 Kiel, Germany*
[3]*Kiel, Nano, Surface, and Interface Science − KiNSIS, Kiel University, 24098 Kiel, Germany, E-mail: talebi@physik.uni-kiel.de*

**Abstract**: Degeneracy is an omnipresent phenomenon in various physical systems, which has its roots in the preservation of geometrical symmetry. In electronic and photonic crystal systems, very often this degeneracy can be broken by virtue of strong interactions between photonic modes of the same energy, where the level repulsion and the hybridization between modes causes the emergence of photonic bandgaps. However, most often this phenomenon does not lead to a complete and inverted bandgap formation over the entire Brillouin zone. Here, by systematically breaking the symmetry of a two-dimensional square photonic crystal, we investigate the formation of Dirac points, line node singularities, and inverted bandgaps. The formation of this complete bandgap is due to the level repulsion between degenerate modes along the line nodes of a semimetal-like photonic crystal, over the entire Brillouin zone. Our numerical experiments are performed by a home-build numerical framework based on a multigrid finite element method. The developed numerical toolbox and our observations pave the way towards designing complete bandgap photonic crystals and exploring the role of symmetry on the optical behaviour of even more complicated orders in photonic crystal systems.

## 1. Introduction

Exploring the role of geometrical symmetry and topology on the electronic and optical responses of solid state systems is an ever-interesting field of physics. Particularly being able to tailor the dispersion of the optical and matter waves via a periodic potential landscape, for example in lattices of atoms or nanostructures, allows for fabulous applications in either integrated photonics [1-4] or electronics systems [5, 6], and provisionally the photonic-electronic interconnects [7-9]. In addition, manipulating the spin, the phase, or the angular momentum of the optical or electronic wave functions provide control knobs for further engineering the energy and the propagation directions of the waves via interactions between the linear and angular momenta [10, 11]. The emerging field of topological photonics is therefore aimed at providing a profound understanding of the role of symmetry and topology on the lattice-wave interactions [12, 13]. Analogous to the discovery of the topological insulators [14, 15] and the role of adiabatic Berry phase in the formation of momentum-space singularities [16, 17], the search for photonic structures that preserve topological effects is ever increasing [18-20]. Normally though, the approach is to mimic the already existing solid-state systems such as the honeycomb lattice of graphene [21-23], or to use magneto-optical materials [24, 25]. Hence, a systematic approach for understanding the role of symmetry – particularly inversion symmetry – in the formation of the complete bandgaps and inverted bands is missing. The latter is of paramount importance for the formation of topologically-protected photonic edge states [26]; nevertheless, it should be accompanied by a complete bandgap so that the emerging surface states are well isolated from the bulk excitations.

Unlike electronic solid-state systems, photonic crystals though offer a higher flexibility in arranging the elements and designing structures of various morphologies and filling ratios. Nanofabrication techniques such as electron-beam lithography [27] and focused-ion milling [28-30] could allow for the realization of such materials with a high accuracy. Such photonic crystal structures could further be utilized in hybrid systems for devising active platforms for further tailoring the band structure by means of strong exciton-plasmon-photon coupling [31]. Hence, a systematic investigation of the role of inversion symmetry on the formation of complete bandgaps paves the way towards the design of photonic crystal structures with on-demand specifications. Nevertheless, a systematic exploration requires also fast numerical schemes that allow for performing parametric sweeping over a large domain of the degrees of freedoms, such as the position and the size of the elements of the crystal.

Here, we systematically explore the role of the inversion symmetry on the formation of the complete bandgap, using a fast numerical scheme based on a home-built multigrid finite-element method. We choose a two-dimensional (2D) square lattice of nanorods with a relatively low permittivity of $\varepsilon_r = 4$ (equal to the permittivity of $Si_3N_4$ in the visible range), and show that a supercell of this structure behaves like a semimetal and supports node line singularities in its reciprocal space [32]. By changing the size and the orientation of the elements, and hence altering the lattice to a lattice with a basis that lacks the inversion symmetry along specific directions or as a whole, we show the optimum conditions where a large complete bandgap is obtained. We further explore the formation of surface waves within the bandgap of the bulk lattice. Our observations could pave the way towards the design and realization of more complicated functionalities based on engineering the symmetry of simple photonic crystal structures.

## 2. Numerical method

Our method is based on Maxwell's equations, particularly the variational formulation

$$\int \frac{1}{\varepsilon(x)} \operatorname{curl} \vec{v}(x) \operatorname{curl} \vec{u}(x) dx = \lambda \int \vec{v}(x) \vec{u}(x) dx \quad (1)$$

for a suitable space of vector-valued test functions $\vec{v}(x)$ with the unknown eigenvector $\vec{u}(x)$ and the unknown eigenvalue $\lambda$, where $\varepsilon(x)$ is the inhomogeneous dielectric permittivity. Since the permittivity is not constant, we cannot use the Helmholtz decomposition to reduce the equation to a scalar-valued problem. We discretize the variational equation using Galerkin's method and choose the space $V_h$ of bilinear Nédélec edge element functions in order to avoid spurious modes in the discrete system [33]. This leads to a matrix eigenvalue problem

$$Ae = \lambda Me \quad (2)$$

with the stiffness matrix $A$ and the mass matrix $M$. Both matrices are self-adjoint, $A$ is positive semidefinite, and $M$ is positive definite. There are various numerical methods available for treating this kind of system, but given that $A$ tends to be fairly large and ill-conditioned, the algorithm has to be chosen carefully. We base our approach on the preconditioned inverse iteration [34]

$$e_{m+1} = e_m - N(Ae_m - \lambda(e_m)Me_m) \quad (3)$$

with the Rayleigh quotient

$$\lambda(e) = \frac{e^T A e}{e^T M e}, \quad (4)$$

where we use a geometric multigrid preconditioner for the matrix $N$. In order to obtain stable convergence, a block Gauß-Seidel method similar to ref. [35] is employed, where each block consists of all edges sharing a common vertex.

Without modification, the inverse iteration will converge to an eigenvector for the smallest eigenvalue, and since the *curl* operator has an infinite-dimensional null space consisting of all gradients of scalar potentials, this is not the desired outcome. We face this challenge by taking advantage of the fact that every gradient *in the Nédélec space* is the gradient of a potential in the space $W_h$ of standard bilinear nodal elements on the same mesh. Following the idea of Hiptmair [36], we employ a lifting mapping a potential $\varphi$ in $W_h$ to its gradient in $V_h$ in combination with a few steps of the geometric multigrid iteration for Poisson's equation to approximately remove gradients from the eigenvector approximations. This leads to a modified inverse iteration

$$e_{m+1} = P(e_m - N(Ae_m - \lambda(e_m)Me_m)), \quad (5)$$

where $P$ denotes the (approximate) orthogonal projection into the space of discrete divergence-free functions.

In order to find band gaps, we have to compute multiple eigenvalues. We do this by applying the preconditioned inverse iteration to multiple eigenvector approximations $e_{m,1}, \cdots, e_{m,k}$ simultaneously. Without modifications, all vectors would converge to the same eigenspace, so we combine the inverse iteration with a Rayleigh-Ritz step: after the approximations for the next step $e_{m+1,1}, \cdots, e_{m+1,k}$ have been computed, we orthonormalize the vectors by solving a projected eigenvalue problem in the subspace they span. This orthonormalization guarantees that we always have a well-defined orthonormal basis of a $k$-dimensional subspace, and that it is a basis of approximate eigenvectors if the inverse iteration is successful.

We stop the inverse iteration if the norm of the residual $|Ae - \lambda(e)Me|$ drops below a given bound, in our case $10^{-2}$ proved to be efficient. This stopping criterion guarantees a sufficient accuracy of the eigenvalue due to the theorem of Bauer et al. [37]. Since the convergence of the $j$-th eigenvector depends on the ratio $\lambda_j/\lambda_{k+1}$, where

$\lambda_1 \leq \lambda_2 \leq \cdots \lambda_n$ are the eigenvalues, it is a good idea to choose $k$ a little larger than the number $m$ of eigenvalues we are actually interested in, since this guarantees that the "interesting" eigenvectors will converge at a rate of at least $\lambda_m/\lambda_{k+1}$. In our experiment, a choice of $k = m + 8$, i.e., the inclusion of eight "throw-away eigenvectors", has led to good results.

The eigenvectors and eigenvalues crucially depend on the chosen boundary conditions. In our case, we use Bloch boundary conditions of the form [38]

$$u_y(a, y) = \exp(ik_x a) u_y(0, y), \qquad (6)$$

and

$$u_x(x, a) = \exp(ik_y a) u_x(x, 0), \qquad (7)$$

for the tangential components of the eigenvector $u$ at the boundaries of the simulation domain, where the Bloch parameter $(k_x, k_y)$ varies in $\left[-\frac{\pi}{a}, \frac{\pi}{a}\right] \times \left[-\frac{\pi}{a}, \frac{\pi}{a}\right]$. Starting the inverse iteration *from scratch* for every Bloch parameter seems wasteful, since the eigenvalues and eigenvectors can be expected to depend continuously (except for special cases) on the matrices $A$ and $M$. To take advantage of this property, we employ extrapolation: if eigenvectors for $(k_x', k_y')$, $(k_x'', k_y'')$, and $(k_x''', k_y''')$ have already been computed, it seems reasonable to expect that the eigenvectors of a nearby parameter $(k_x, k_y)$ can be approximated by the previous eigenvectors. Instead of applying standard polynomial extrapolation, we take advantage of the Courant/Fischer [39, 40] theorem that states that, since $A$ and $M$ are self-adjoint, the eigenvalues are the local minima of the Rayleigh quotient $\lambda(e)$: we look for minima of the Rayleigh quotient in the subspace spanned by the eigenvectors corresponding to three previous Bloch parameters. This is equivalent to solving a $3k$-dimensional eigenvalue problem, and this approach allows us to reduce the number of iterations to two or three for most of the Bloch parameters.

To benchmark the accuracy of our numerical method, we have calculated the dispersion diagram of a simple square lattice of holes composed of circles with the permittivity of $\varepsilon_r = 11.56$ and compared them with the results published elsewhere [41]. As shown in the supplementary Fig. S1, we observe a full agreement.
Compared to several methods for numerical electromagnetism that could be applied to compute the band structure of photonic crystals, such as finite-difference time-domain method [42] and plane wave expansion method [43] our method has the advantage of being more efficient, particularly for meshes with high resolutions. The high efficiency results from a combination of multiple techniques: (i) Since a regular mesh is employed, there is no need to store the matrix coefficients explicitly, but they can be computed on the fly as needed. This reduces the load on the computer's memory bus and allows us to take better advantage of cache memory. (ii) The multigrid method with block smoothing provides us with a good preconditioner that requires only a few sweeps across the mesh and no large amount of additional memory, thus allowing us to handle meshes with millions of degrees of freedom on standard workstations. (iii) The preconditioned inverse iteration also requires only a few sweeps across the mesh to compute linear combinations and dot products, and using the well-tested and reliable LAPACK routines to solve the projected problems in the subspaces lets us take advantage of decades of expertise in traditional eigenvalue solvers. (iv) All operations on the grid functions have been parallelized using the OpenMP standard, only the low-dimensional eigenvalue problems in the subspaces and the coarse-grid computation are performed by only one core of the processor, since they do not offer a significant amount of concurrency. (v) Like any non-linear iteration, the preconditioned inverse iteration crucially depends on good starting values, and here the role of our extrapolation technique cannot be overstated: if we just use the previous eigenvectors, we require seven iterations to arrive at a suitable accuracy, while with second-order extrapolation two iterations suffice, and the run-time is reduced by more than 70%. (vi) The speed of convergence can be improved significantly by computing "throw-away eigenvectors" that only serve to increase the gap between the relevant eigenvalues and the remainder of the spectrum. Even on a simple workstation with a 12-core desktop processor, computing eigenvectors for the 16 lowest non-zero eigenvalues requires less than 30 seconds for more than 2 million degrees of freedom once the extrapolation takes hold after the first few steps.

## 3. Results and discussions

### 3.1. Bulk response

We consider first a highly symmetric system composed of a supercell of a square lattice of rods with the radius of 0.125 $a$, and the permittivity of $\varepsilon_r = 4$ where $a$ is the distance shown in Fig. 1a. The permittivity of the lattice elements considered here is considerably lower than the structures discussed in ref. [44]. This allows for a rather easier practical realization, for example with the $Si_3N_4$ material within the visible range. Each unit cell of the supercell thus is composed of 4 irreducible unit cells of the square lattice. The supercell approach allows for a better identification of the degenerate points in photonic crystals with point symmetry [45]. The energy−wave-vector dispersion relation of this structure is calculated using the method mentioned above in the 2D space (Fig. 1b). The peculiar appearance of twin Dirac points in the Brillouin zone of the supercell, leads us to further investigate the origins of this degeneracy. First, the degeneracy and the appearance of the Dirac-point singularities in the dispersion diagram is the consequence of the inversion of the higher energy bands, causing a nodal line singularity, as opposed to either pure Dirac-like dispersion or Weyl points [32]. Particularly, within the momentum range $k_x = 0$ to $0.7\frac{\pi}{a}$, the lower and upper energy bands sustain a dipolar and quadruplar excitations, whereas after the band inversion the mode profiles are reversed (See Fig. S2 for more details regarding the band inversion). This behavior is fully captured by the computed 3D dispersion diagram and could be hardly revealed from a 2D dispersion diagram (Compare Fig. 1b to Fig. 2b). Second, the degeneracy in our structure is due to the coexistence of the optical modes with different spatial profiles and the same frequency. The degeneracy is an omnipresent phenomenon in various optical systems, and could be compared to the degeneracy of the $p$-orbitals or the $d$-orbitals of electrons in atoms. In addition, acoustic topological systems that constitutes acoustic waves propagating on a honeycomb like lattice constitutes similar pseudospin structures as well [46]. The widely discussed topological photonic crystal composed of the triangular lattices of honeycomb configurations[44], uses 6 dielectric rods to form $p_\pm$ and $d_\pm$ orbitals. There, $p_x$ ( $p_y$ ) orbitals are formed by the dipolar resonances sustained by the unit cell of the lattice and they are naturally degenerate. This fact holds also for degenerate quadruple resonances that holds resonances resembling $d_{x^2-y^2}$ and $d_{xy}$ orbitals. Two pseudospin states of the lattices are superpositions of the aforementioned orbitals as $p_\pm = (p_x \pm i p_y)/\sqrt{2}$ and $d_\pm = (d_{x^2-y^2} \pm i d_{xy})/\sqrt{2}$ . In general, though, as will be discussed and demonstrated in supplementary Fig. S2, these forms of optical modes can be manifested by the higher-frequency quadruple modes of a single dielectric disc. Third, the degeneracy of the optical modes is broken by changing the symmetry of the lattice. Particularly, we break the inversion symmetry of the lattice along the *x* and *y* directions, whereas the symmetry along the diagonal is retained. In this way, a simple squared lattice is changed into a centered square lattice. As a result a complete bandgap is formed, where the largest level-repulsion is observed along the previously degenerate Brillouin zone points. Adding this effect of level repulsion to the band-inversion, allow the formation of vortex edge modes as will be discussed later. Nevertheless, since the system is 2D, the transverse vorticity and its locking to the momentum of the edge mods, result in the unidirectional propagation of the optical modes with their pseudospin being locked to their direction of propagation.

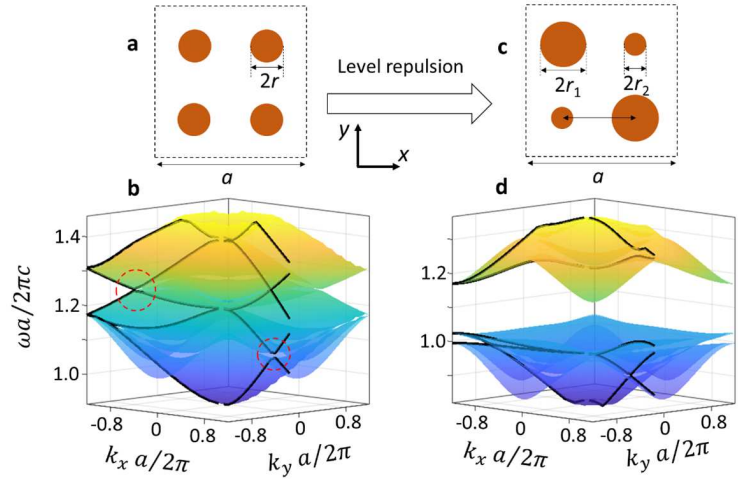

**Fig. 1. Controlling Dirac points and nodal-line singularities in a photonic crystal via symmetry.** By changing the symmetry from a simple square lattice (a) to a centred square lattice (c), nodal line singularities in the reciprocal space (b) are turned into an inverted bandgap (d). Here, $r = 0.125a$, $r_1 = 0.2a$ and $r_2 = 0.1a$. A multimedia representing the change of the two-dimensional band structure, via the systematic change of the ratio $r_1/r_2$ is provided in the Visualization 1. Visualization 2 represents the 3D band structure of the photonic crystal in panel c, step by step.

Noticeably, the inverted bands happening in our supercell of the square lattice (Fig. 1a) is reminiscence of the conduction- and valence-band inversions in various types of topological materials such as $Bi_2Se_3$ and $Bi_2Te_3$. However, the mere existence of band inversion does not allow for topological effects; for the latter to happen, the degeneracy of the bands should be broken, via e.g., spin-orbit interactions, time-reversal symmetry, or chiral symmetry. As we will discuss further in the next section, the lifting up of the degeneracy in our structure is due to the presence of chiral symmetry, and can be qualitatively described using a hopping model.

Before discussing the emergence of the level repulsion and the formation of edge modes, we further investigate the role of point symmetry and the size of the rods on the degeneracy of the optical modes. We present in the following the 2D dispersion diagrams of the investigated systems (Fig. 2). In addition to the photonic crystals already discussed above (Fig. 1 and 2a and b), we consider four other configurations where the point symmetry is systematically altered along a certain direction, by either keeping the sizes of the elements all the same (Fig. 2c and d), or considering two different sizes for the elements (Fig. 2e and f). The simple square lattice itself sustains a complete bandgap between the 4[th] and 5[th] bands, and interestingly, for all lattices with the similar sizes of the rods, the full bandgap is retained, no matter the symmetry of the complete structure. Moreover, for the lattices with different sizes of the elements, again, the complete bandgap is observed between the 8[th] and 9[th] band, and only the size of the gap is slightly altered. This leads us to the conclusion that the topological behavior of the system is already linked to the optical modes and resonances of individual elements. At $\omega a/2\pi c = 0.5$, the dispersion diagram of the structure shown in Fig. 2a already shows a specific Dirac-point degeneracy at the M point, very similar to the behavior of the Dirac-like dispersion observed for the triangular lattices of honeycomb structures at the Γ point. As discussed above, this is due to the line node singularity of the inverted bands. In the case of the centered-square lattice (Fig. 2b), the degeneracy is released at the M point, nevertheless, only a local gap is opened. In contrast, the level repulsion along the previously-discussed higher energy points opens a global gap that is highlighted by the gray area. To better describe the lifting up of the degeneracy in structure shown in Fig. 2b, and discuss the band inversion, we have provided a multimedia (Visualization 1), where the radii of the discs are altered gradually (see Supplementary Fig. S2 and the discussions therein).

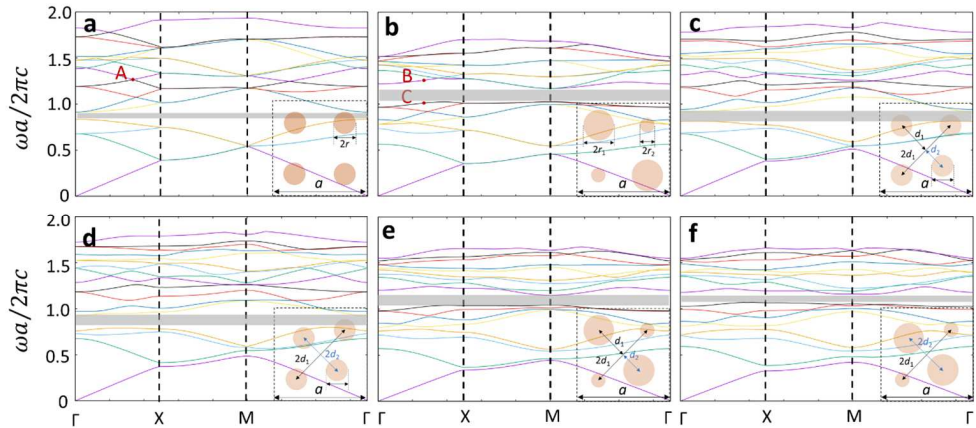

**Fig. 2. Braking the degeneracy of the modes via controlling the symmetry.** Dispersion diagrams of the photonics modes for (a) a simple square lattice, (b) a centred square lattice, (c) a lattice with the similar sizes of rods and a single symmetry axis, (d) a lattice with the similar sizes of rods and two symmetry axes, (e) a lattice with two different sizes of the holes and a single inversion symmetry axis, and (f) a lattice with two sizes of the rods and two symmetry axes.

To explore the spatial profile of the degenerate modes involved in the level repulsion, we calculate the eigen vectors of the highly degenerate Dirac points at the point A marked in Fig. 2a (points A1 and A2 in Fig. 3a and b; calculated at an infinitesimal distance with respect to the point A in Fig. 2a). The amplitudes of both modes show a uniform maximum intensity inside the rods, though the in-plane vector representation of the magnetic fields unravel a completely different symmetry: While the excitation of $d_\pm$ orbitals are apparent (highlighted by green curved arrows), they are locally displaced in different parts of the unit cells for A1 and A2 modes. Moreover, a peculiar vortex field, and a monopole-like singularity are also observed for A1 and A2 modes, respectively, as marked by the red arrow lines. Despite the fact that the vector representation of the magnetic field resembles the excitation of induced magnetic monopoles, the zero intensity of the magnetic field at those points rules out the existence of magnetic monopoles. By changing the symmetry of the lattice from a simple square lattice to the centered square lattice, the hybridized modes show a mixed vortex and monopole-like singularity (Fig. 3, point C and D). Moreover, they sustain symmetric and antisymmetric profiles inside the rods along the y direction, for the symmetric mode occupying a lower energy state.

*3.2. Edge modes:*

Having the underlying topological aspects, we further explore the dispersion of possible surface states inside the bandgap. For this purpose, we use a supercell approach, where the structure remains periodic along the *x* axis and we have considered 12 unit cells for the length of the structure along the *y* axis. Moreover, the structure is slightly altered, so that the elements are slightly squeezed inside the unit cell, such that the inter-cell distances between rods (considered to be $0.1a$) are smaller than the intra-cell distances (being equal to $0.3a$). In this way, the symmetry between intra-cell and inter-cell hopping amplitudes are broken and as a result of the emerging chiral symmetry, nontrivial bandgap and topological effects become apparent.

COMSOL multiphysics is used for exploring the dispersion of the vortex edge modes. Two surface modes are obtained that remain degenerate for the wavenumber range within $k_x = \left[0, 0.36\ \pi a^{-1}\right]$ and further form two distinguished modes beyond this range. Interestingly, this is exactly the point where the line node singularities are gapped, by introducing the asymmetry in the crystal. The field profile of the demonstrated modes are retained bound to the first two layers of the lattice and propagate via a zigzag-like coupling between the lattice elements of the first two rows (see Visualizations 3 and 4). Moreover, because the propagation phase of the edge waves are larger than the free-space wavenumber, the edge modes are bound to the interface between the lattice and free space, without leaking into the free space, in contrast with lattices where the band inversion and gaps are centered at the *Γ* point of the Brillouin zone. In the latter cases, an interface between two photonic crystals of trivial and nontrivial bandgaps is to be considered for the realization of bound modes [19, 46, 47]. The leakage-free behavior of the edge modes in the present system allows for a simpler coupling of the incident field of emitters like scanning near-field tips and electron beams and easier experimental characterization. In general, two edge modes are observed, with reversed direction of pseudospin and propagation direction (Fig. 4c). This leads to the conclusion that the direction of the linear momentum and the vorticity of the magnetic field (pseudospin of the field) are indeed locked and reversed for A and B modes as shown in Fig. 4c. Moreover, it should be noticed that evanescent waves in general sustain a transverse spin. The propagation direction of surface plasmon polaritons is locked to their transverse spin [48]. However, normally electric-field polarization is explored for the spin-momentum locking aspects; in contrast here, the vorticity of the magnetic field (the pseudospin direction) is locked to the propagation direction of the edge mode.

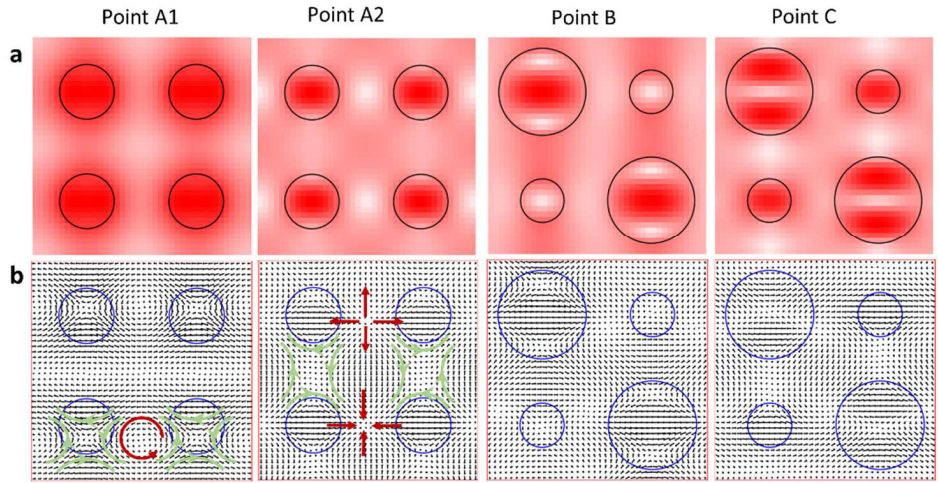

**Fig. 3. In-plane magnetic field profiles of the optical modes at the points depicted in Figure 3, for the first and second structure.** The degenerate optical modes with their *k*-vectors located along the nodal lines, represent two modes with vortex (A1) and monopole-like (A2) symmetries in the real space. The spatial profiles of the absolute value of the total magnetic field and the vector representations are shown in (a) and (b) respectively. By changing the symmetry from a simple square lattice to a centred square lattice, two non-degenerate modes are formed as a result of level repulsion (hybridization) (B and C; marked in Fig. 2), having mixed vortex- and monopole-like characteristics. Color scale changes from 0 (white) to 1 (red).

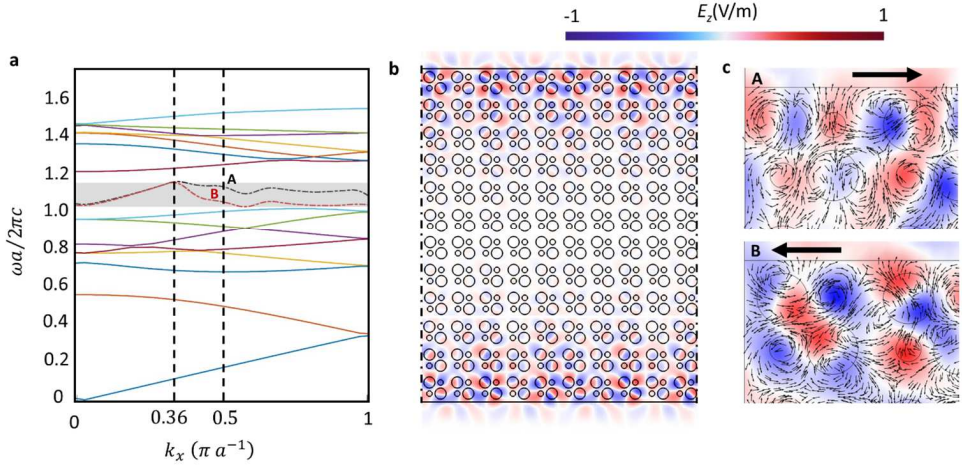

**Fig. 4. Bound surface modes.** (a) Calculated dispersion diagram of the supercell along the *x* direction, where 12 unit cells where considered along the *y*-axis. (b) $E_z$ Field profile of the A mode within the supercell at $k_x = 0.5\pi/a$, and (c) the quiver motion of the in-plane magnetic field (vector field) and the amplitude of the *z*-component of the electric field presented in a smaller region for a better clarity. Black arrows show the direction of propagation of each mode. Visualizations 3 and 4 show the dynamics of the edge modes A and B, respectively.

### 3.3. Chiral symmetry and generalized SSH model

Topological photonic structures can be categorized according to their symmetries, into (i) structures with broken time-reversal symmetries [25], (ii) quantum spin-hall structures with spin-orbit interactions underpinning the topological effects [44, 46], and (iii) structures with chiral symmetry [48, 49]. For the latter case, typically the one-dimensional Su-Schrieffer-Heeger (SSH) model is adopted on a bipartite grid [50]. Thus, a chain of coupled nanoparticles could be used to practically realize the SSH Hamiltonian. In the followings, we show that the behavior of the photonic crystal structure explored above, is qualitatively describable using a generalized SSH model, by adopting a hopping scheme for the photons on the lattice.

For simplicity, we consider a one-dimensional chain, with three inter-cell hopping amplitudes, as shown in Fig. 5a. The intra-cell hopping amplitudes are the same as the inter-cell ones. According to the one-dimensional SSH model, there exists no gap for such a case; however, we show here that the interplay between three hopping amplitudes allows for line node singularities and further breaking their degeneracy as observed for our topological photonic structure. The single-particle Hamiltonian of the structure, can be written as

$$\begin{aligned}
\hat{H} = &t_1 \sum_{m=1}^{M} \{|m,A\rangle\langle m,B| + |m,A\rangle\langle m,C| + |m,B\rangle\langle m,D| + |m,C\rangle\langle m,D|\} \\
&+ t_2 \sum_{m=1}^{M} |m,A\rangle\langle m,D| + t_3 \sum_{m=1}^{M} |m,B\rangle\langle m,C| + t_1 \sum_{m=1}^{M-1} \{|m,B\rangle\langle m+1,A| + |m,D\rangle\langle m+1,C|\} \\
&+ t_2 \sum_{m=1}^{M-1} |m,D\rangle\langle m+1,A| + t_3 \sum_{m=1}^{M-1} |m,B\rangle\langle m+1,C| + \text{h.c.},
\end{aligned} \quad (8)$$

where $|m,\alpha\rangle$ with $\alpha \in \{A,B,C,D\}$ denoting the state of the chain where the photon is on the element site $\alpha$ on the $m^{\text{th}}$ unit cell, and h.c. stands for the Hermitian conjugate. $M$ is the total number of the chain elements.

First, we consider only the bulk response by assuming the periodic Born-von-Karman boundary conditions, taking the periodicity in the Brillouin zone as $\hat{H}(k+2\pi) = \hat{H}(k)$ and $\hat{H}_{\text{bulk}}|\alpha(k)\rangle = E(k)|\alpha(k)\rangle$, we obtain

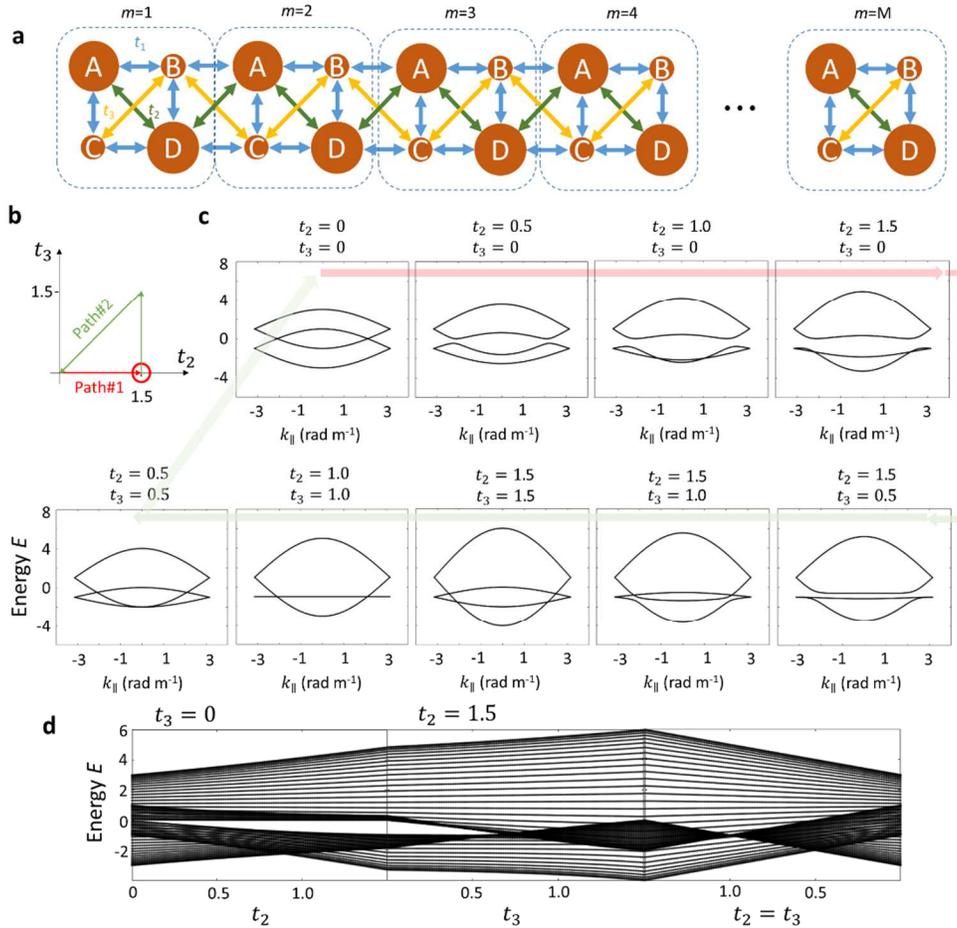

**Fig. 5. Hopping model on the bipartite network, where the intercell and intracell hopping amplitudes are equal.** (a) Geometry of the generalized SSH model, including three different hopping amplitudes. (b) Paths taken on the $(t_2, t_3)$ coordinates for $t_1 = 1$, where the red path opens the gap in the band structure, and the green path closes the gap. (c) calculated bandstructure of the bulk lattice on the path depicted in panel (b), by assuming the Born-von-Karman boundary conditions. (d) The energy diagram of a truncated lattice, with 15 unit cells, along the path shown in panel b.

$$\begin{bmatrix} 0 & t_1(1+e^{-ik}) & t_1 & t_2(1+e^{-ik}) \\ t_1(1+e^{ik}) & 0 & t_3(1+e^{ik}) & t_1 \\ t_1 & t_3(1+e^{-ik}) & 0 & t_1(1+e^{-ik}) \\ t_2(1+e^{ik}) & t_1 & t_1(1+e^{ik}) & 0 \end{bmatrix} \begin{bmatrix} a(k) \\ b(k) \\ c(k) \\ d(k) \end{bmatrix} = E(k) \begin{bmatrix} a(k) \\ b(k) \\ c(k) \\ d(k) \end{bmatrix}, \quad (9)$$

where $a(k)$ to $d(k)$ are the Fourier-space amplitudes of the wave functions on the sub-elements A to D. Equation (9) is solved to obtain the eigen energies and eigen states of the system at each given $k$ within the range $(-\pi, \pi)$. Noticeably, this matrix cannot be simplified to highlight the pseudospin symmetries via Pauli matrices, as our system sustains 4 eigen states dissimilar to the SSH model that constitutes a two-level system. Nevertheless, still one can derive a concise form using the Weyl representation of Dirac's matrices. Thus, eq. (9) is recast as

$$\hat{H} = \left\{ -t_1 \gamma_0 \gamma_5 \gamma_1 + \frac{t_2+t_3}{2} \gamma_1 \gamma_5 + i \frac{t_2-t_3}{2} \gamma_2 \right\}(1+\cos k)$$
$$+ \left\{ -t_1 \gamma_0 \gamma_5 \gamma_2 + \frac{t_2+t_3}{2} \gamma_2 \gamma_5 - i \frac{t_2-t_3}{2} \gamma_1 \right\} \sin k \quad (10)$$
$$+ t_1 \gamma_0$$

with

$$\gamma_0 = \begin{pmatrix} 0 & 1_2 \\ 1_2 & 0 \end{pmatrix}, \quad (11)$$

$$\gamma_i = \begin{pmatrix} 0 & \hat{\sigma}_i \\ -\hat{\sigma}_i & 0 \end{pmatrix}, \quad (12)$$

and $\gamma_5 = i\gamma_0\gamma_1\gamma_2\gamma_3$, where $\hat{\sigma}_i$ are the Pauli matrices.

The topological effects in the present structure is affected by the interplay between three hopping amplitudes $t_1$, $t_2$, and $t_3$. Obviously, $t_1 = 0$ decouples the bipartite lattice into two individual and uncoupled grids. Thus, we assume $t_1 = 1$ and calculate the band structure on a closed path in the $(t_2, t_3)$ space, shown in Fig. 5b. For $t_1 = 1$, $t_2 = 0$, $t_3 = 0$, the band structure does not experience a gap, though two merging bands, mimicking the line node singularity of the photonic crystal structure investigated above is observed (Fig. 5c). By taking the path $t_2 = 0 \to 1.5$ and $t_3 = 0$, the band structure undergoes a global gap formation, and by increasing the value of $t_2$ from $t_2 = 0.5$ to $t_2 = 1.5$, the first and second bands are inverted. Now by taking the path along the green arrows shown in Fig. 5b and coming back to the origin of the $(t_2, t_3)$ space, the gap is again closed. To investigate whether the geometrical lattice explored here could support and edge mode, the eigenstates and eigen energies of a truncated lattice is investigated, by applying the Hamiltonian in eq. (8) to the wavefunctions $\{|A\rangle, |B\rangle, |C\rangle, |D\rangle\} \otimes (1:M)$. The results for $M = 15$ is shown in Fig. 5d. Obviously, the structure does not show any edge mode. Hence, the opened gap should be a trivial gap. This is consistent with the one dimensional SSH model with equal inter-cell and intra-cell hopping amplitudes.

To observe topological effects and edge modes, the hopping amplitudes between the inter- and intra-cell elements must differ. Thus, we generalize the geometrical lattice by adopting 6 different hopping amplitudes, shown in Fig. 6a. In this way, eq. (9) is transferred to

$$\begin{bmatrix} 0 & t_1 + t_1' e^{-ik} & t_1 & t_2 + t_2' e^{-ik} \\ t_1 + t_1' e^{+ik} & 0 & t_3 + t_3' e^{+ik} & t_1 \\ t_1 & t_3 + t_3' e^{-ik} & 0 & t_1 + t_1' e^{-ik} \\ t_2 + t_2' e^{+ik} & t_1 & t_1 + t_1' e^{+ik} & 0 \end{bmatrix} \begin{bmatrix} a(k) \\ b(k) \\ c(k) \\ d(k) \end{bmatrix} = E(k) \begin{bmatrix} a(k) \\ b(k) \\ c(k) \\ d(k) \end{bmatrix}, \quad (12)$$

where the matrix remains h.c.

When the inter-cell values are different from intra-cell hopping amplitudes, another gap is opened, though this time at the boundaries of the Brillouin zone. For inter-cell hopping amplitudes being smaller than the intra-cell ones, a trivial gap is formed, whereas for the case when the inter-cell hopping amplitudes are larger, a nontrivial gap is formed (Fig. 6b). This can be confirmed by the calculating the dispersion of the edge modes, by considering a truncated lattice for example (Fig. 6c). Thus whereas in both cases. i.e., with equal inter- and intra-cell hopping amplitudes and for different cases, a global gap could be achieved, only the latter lattice could show nontrivial gap formations and support topological edge effects.

The winding number associated to the generalized SSH model here, is the summation of the winding numbers allocated to the upper half elements above the diagonal of the matrix presented in eq. (12). In this way, since the winding number associated to each element is nonzero, the total winding number is also nonzero.

Our model thus reproduces the formation of the line-node singularities and the associated level repulsion by introducing asymmetric intra-cell hopping amplitudes, as well as the formation of nontrivial gaps via an imbalance between the inter-cell and intra-cell hopping amplitudes. Due to the presence of chiral symmetry, the hopping model could better systematically describe the lifting up of the degeneracy at the line-node singularities, compared to the $\vec{k} \cdot \vec{p}$ method [52].

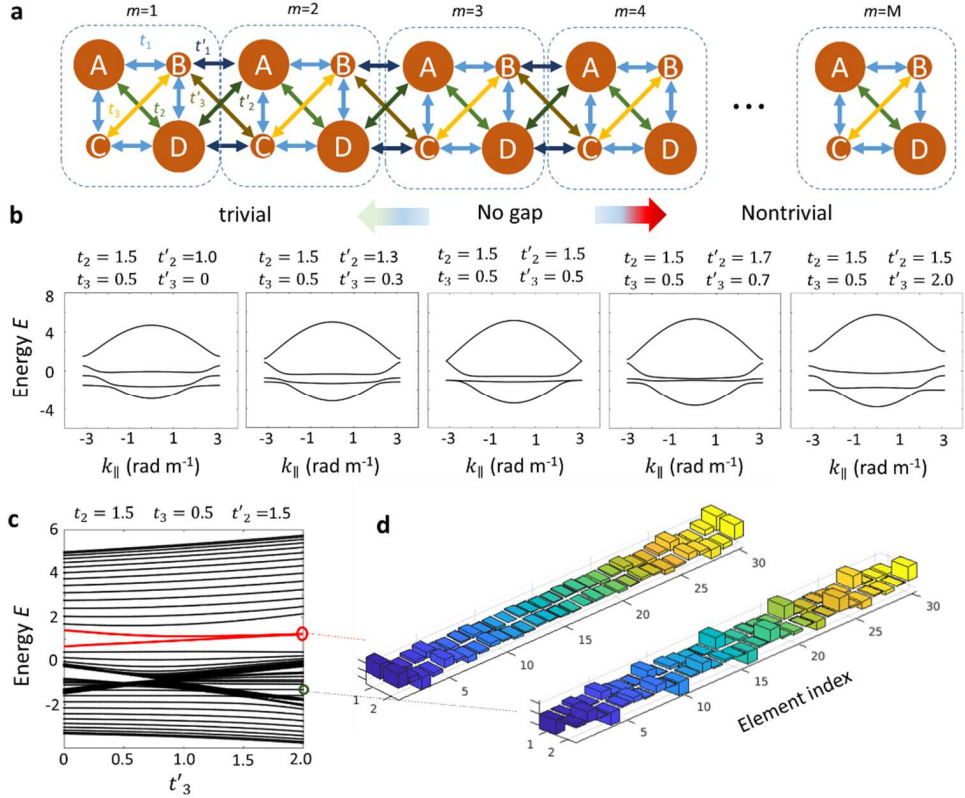

**Fig. 6. Chiral symmetry and nontrivial gap formation.** (a) Geometrical SSH model with 6 different hopping amplitudes for the inter-cell and intra-cell couplings. (b) The calculated band structure along the path leading to the first trivial gap formation for $t'_2 < t_2$ and $t'_3 < t_3$, and second a nontrivial gap formation for $t'_2 > t_2$ and $t'_3 > t_3$. (c) Calculated energy states for a truncated lattice with $M = 15$ elements versus different $t'_3$ values and for depicted values for other hopping amplitudes. (d) The eigen states of the lattice for two depicted eigen values, denoting an edge state (top) and a bulk state (bottom). For all cases, $t_1 = t'_1 = 1$.

The hopping model described here is a general scheme to qualitatively show the formation of nontrivial chiral symmetry in our structure. Nevertheless, the practical realism of the present model relies on schemes to tailor the coupling efficiencies systematically, as explored before here. We notice though, that not always bringing two elements nearer to each other underpins a stronger coupling and larger hopping amplitudes. Particularly, for our case, the opposite case is valid, where we indeed notice that the edge modes could be obtained by further increasing the inter-cell distance between elements. We allocated this behavior to the competition between near-field and far-field multipolar couplings and energy transfer, hence an optimistic situation is to be explored using a numerical scheme as explored here.

## 4. Conclusion

Line node singularities in a semimetal-like photonic crystals are demonstrated within a supercell of a simple square lattice, composed of 2D rods with the permittivity of 4. The degeneracy of the optical modes along the nodal lines are broken by a systematic introduction of asymmetry, where the symmetry is altered along the x and y directions. Since the merging bands of the initial lattice are by definition inverted – since the reciprocal lattice sustain a line node singularity – topological aspects and the formation of surface modes in the symmetry-broken crystal is an expected aspect. Further, a global bandgap is obtained with the optical modes that inherit the pseudospin characteristics of the initial highly symmetric photonic crystal. These pseudospin characteristics allow for the emergence of two surface modes that are further fully characterised and discussed. Our observations lead us to the conclusion that topological aspects are highly expected to happen even in simpler crystals at higher energies, where the superposition of quadrupole resonance lead to the emergence of *d*-like electronic orbitals. Moreover, considering a much lower permittivity for the material, this system could be fabricated and characterised at optical frequencies. For example, $Si_3N_3$, due to its low dissipation at optical ranges is a promising candidate for realizing this topological photonic structure. In addition, high-resolution characterization techniques, such as cathodoluminescence spectroscopy and scanning near-field optical microscopy could be used to directly resolve the edge modes and their vorticity when combined with polarimetry techniques.

**Funding.**

This project received financial support from the European Research Council (ERC) under the European Union's Horizon 2020 research and innovation programme, Grant Agreements No. 802130 (Kiel, NanoBeam) and Grant Agreements No. 101017720 (EBEAM).

**Acknowledgement.**

All authors gratefully thank Anand Srivastav (Kiel University) for fruitful discussions and Robin Lautenbacher for recasting equation (9) in terms of Dirac's matrices. All authors acknowledge support from KiNSIS Networked Matter Challenge initiative.

**Disclosures.**

The authors declare no conflicts of interest.

**Data availability.**

Data underlying the results presented in this paper are not publicly available at this time but may be obtained from the authors upon reasonable request.

**Supplemental document.**

See Supplement 1 for supporting content.